\documentclass[twocolumn,aps,prl]{revtex4}
\usepackage{graphicx}
\usepackage{times}

\begin{document}

\title{Gravitational Baryogenesis}

\author{Hooman Davoudiasl}
\author{Ryuichiro Kitano}
\author{Graham D. Kribs}
\author{Hitoshi Murayama}
\thanks{On leave of absence from Department of Physics,
University of California, Berkeley, CA 94720.}
\author{Paul J. Steinhardt}
\thanks{On leave of absence from Department of Physics, Princeton
University, Princeton, NJ 08544.}
\affiliation{School of Natural Sciences, Institute for Advanced
Study, Princeton, NJ 08540}

\date{March 1, 2004}


\begin{abstract}
  We show that a gravitational interaction between the derivative of
  the Ricci scalar curvature and the baryon-number current dynamically
  breaks $CPT$ in an expanding universe and, combined with
  baryon-number-violating interactions, can drive the universe towards
  an equilibrium baryon asymmetry that is observationally acceptable.
\end{abstract}
\maketitle


The successful predictions of Big-Bang Nucleosynthesis (BBN)
\cite{Burles:2000ju}, highly precise measurements of the cosmic
microwave background \cite{WMAP}, and the absence of intense radiation
from matter-antimatter annihilation \cite{CGdeR} all indicate that the
universe contains an excess of matter over antimatter.  Numerically,
the baryon-to-entropy ratio is $n_B/s = 9.2^{+0.6}_{-0.4} \, \times
10^{-11}$.  What remains a mystery is how the baryon asymmetry was
generated.

The contemporary view is that the baryon asymmetry is generated
dynamically as the universe expands and cools.  Sakharov
\cite{Sakharov} argued that three conditions are necessary: (1)
baryon-number non-conserving interactions; (2) $C$ and $CP$ violation;
and, (3) a departure from thermal equilibrium.  To satisfy the latter
two conditions, the conventional approach has been to introduce
interactions that violate $C$ and $CP$ {\it in vacuo} and a period
when the universe is out of thermal equilibrium.

In this paper, we propose a mechanism that generates an
observationally acceptable baryon asymmetry while maintaining
thermal equilibrium.  The key ingredient is a $CP$--violating
interaction between the derivative of the Ricci scalar curvature
${\cal R}$ and the baryon number ($B$) current $J^{\mu}$:
\begin{equation} \label{eq1}
  \frac{1}{M_*^2} \int d^4 x \sqrt{-g} (\partial_{\mu} {\cal
    R})J^{\mu},
  \label{eq:operator}
\end{equation}
where $M_*$ is the cutoff scale of the effective theory.  It is not
necessary that $J^{\mu}$ be the $B$ current; any current that leads to
net $B-L$ charge in equilibrium ($L$ is the lepton number), so that
the asymmetry will not be wiped out by the electroweak anomaly
\cite{Kuzmin:1985mm}, is sufficient for our purpose.  It is natural to
expect such an operator in the low-energy effective field theory of
quantum gravity if $M_*$ is of order the reduced Planck scale $M_P =
(8\pi G_N)^{-1/2} \simeq 2.4\times 10^{18}$~GeV.  We also note that it
can be obtained in supergravity theories from a higher dimensional
operator in the K\"{a}hler potential.

The interaction in Eq.~(\ref{eq:operator}) violates $CP$ and is $CPT$
conserving {\it in vacuo}\/. However, this interaction dynamically
breaks $CPT$ in an expanding universe and biases the energetics in
favor of causing an asymmetry between particles and anti-particles.

To generate a baryon asymmetry using the interaction in
(\ref{eq:operator}), we also require that there be $B$-violating
processes in thermal equilibrium.  We denote the temperature at which
$B$-violation decouples by $T_D$.  Given these ingredients, the $B$
asymmetry in our setup is generated as follows. In an expanding
universe, where ${\cal R} \sim H^2$ and $\dot{{\cal R}}$ are non-zero
(where a dot means time derivative), the interaction in
Eq.~(\ref{eq1}) gives opposite sign energy contributions that differ
for particle versus antiparticle, and thereby dynamically violates
$CPT$. This modifies thermal equilibrium distributions in a similar
fashion as a chemical potential $\mu \sim \pm \dot{{\cal R}} /M_*^2$,
driving the universe towards non-zero {\it equilibrium} $B$ asymmetry
via the $B$-violating interactions.  Once the temperature drops below
$T_D$, as the universe expands and cools, a net asymmetry remains:
\begin{equation}
  \frac{n_B}{s} \approx \left.\frac{\dot{{\cal R}}}{M_*^2
      T}\right|_{T_D}. \label{eq:master}
\end{equation}

Our approach is closely related to ``spontaneous baryogenesis''
\cite{Cohen:1987vi}, which relies on the derivative coupling
between a spatially uniform scalar field and the baryon number
current, $(\partial_{\mu} \varphi ) J^{\mu}$.  With $\varphi$,
though, the construction is considerably more complicated.  The
scalar, essential to this mechanism, has to be added by hand,
whereas the term in Eq.~(\ref{eq:operator}) is expected to be
present in an effective theory of gravity. The initial conditions
for $\varphi$ must be separately specified and justified:
$\varphi$ must be forced to evolve homogeneously in one direction
versus the other to produce an asymmetry and must be spatially
uniform. In contrast, the time-evolution of ${\cal R} \propto H^2$
is required in a cosmological background and it is highly
spatially uniform because the universe is highly homogeneous.  The
oscillation of $\varphi$ around its minimum is also a complication
because the average $\dot{\varphi}$ is zero, tending to cancel the
asymmetry \cite{Dolgov:1996qq}, whereas the mean value of
$\dot{\cal R} \sim H^3$ does not vanish.

To produce baryon asymmetry by the gravitational interaction in
Eq.~(\ref{eq1}), several factors have to be considered.  For a
constant equation of state $w$, where $w$ is the ratio of the
pressure $p$ to the energy density $\rho$, the scalar curvature is
proportional to $(1-3w)$, and its time derivative is given by
\begin{equation}
  \label{eq:1}
  \dot{\cal R} = -(1-3w)\frac{\dot{\rho}}{M_P^2} = \sqrt{3}
  \, (1-3w)(1+w)\frac{\rho^{3/2}}{M_P^3}.
\end{equation}
We will examine $B$ generation for some cosmologically important
values of $w$.

We start from the radiation-dominated era following inflation when $w
\approx 1/3$.  If $w$ were equal to precisely 1/3, then the right-hand
side of Eq.~(\ref{eq:1}) would vanish and there would be no effect.
However, $w=1/3$ only applies in the limit of exact conformal
invariance, $T_{\mu}^{\mu} =0$.  In practice, interactions among
massless particles lead to running coupling constants, and, hence, the
trace anomaly that makes $T_{\mu}^{\mu} \propto \beta(g) F^{\mu \nu}
F_{\mu \nu}\neq 0$.  The thermodynamic potential of a plasma of the
SU($N_c$) gauge theory, with coupling $g$ and $N_f$ flavors, has been
worked out in detail \cite{Kajantie:2002wa}.  It leads to
\begin{equation}
  \label{eq:6}
  1-3w = \frac{5}{6\pi^2} \frac{g^4}{(4\pi)^2}
  \frac{(N_c+\frac{5}{4}N_f)(\frac{11}{3}N_c -
    \frac{2}{3}N_f) }
  {2+\frac{7}{2}[N_c N_f/(N_c^2-1)]}
\end{equation}
up to $O(g^5)$ corrections, where the last factor in the numerator is
the beta function coefficient.  Typical gauge groups and matter
content at very high energies can easily yield $1-3w \sim
10^{-2}$--$10^{-1}$.  There may also be mass thresholds that lead to
conformal violation.  Then, Eq.~(\ref{eq:master}) gives
\begin{equation}
  \label{eq:w=1/3}
  \frac{n_B}{s} \approx (1-3w) \frac{T_D^5}{M_\ast^2 M_P^3}.
\end{equation}
The upper bound on tensor mode fluctuations constrains the
inflationary scale to be $M_I \leq 3.3 \times 10^{16}$~GeV
\cite{Peiris:2003ff}, and obviously for this scenario $T_D < T_{RD} <
M_I$, where $T_{RD}$ is the temperature at which the universe becomes
radiation dominated ({\it i.e.}\/, the reheat temperature in this
case). It is remarkable that the asymmetry can be sufficiently large
even for $M_\ast \simeq M_P$ if $T_D \simeq M_I$. In this case, the
scenario predicts that tensor mode fluctuations should soon be
observed.

A second case of cosmic relevance is $w = 0$.  This corresponds to the
matter domination epoch which characterizes, for example, conventional
perturbative reheating via a scalar field $\phi_{osc}$, as it
oscillates around the minimum of a quadratic potential. During the
oscillation phase, $\rho \propto a^{-3}$, $a \propto t^{2/3}$, and
$\phi_{osc}$ decays at a rate $\Gamma$ into radiation, whose energy
density becomes equal to that of the scalar field when $H \simeq
\Gamma \simeq T_{RD}^2/M_P$.  Therefore,
\begin{eqnarray}
  \rho_{osc} &\simeq& T_{RD}^4 \left( \frac{a_{RD}}{a} \right)^3, \\
  \rho_R &\simeq& T_{RD}^4 \left( \frac{a_{RD}}{a} \right)^{3/2}.
\end{eqnarray}
The latter equation suggests $T \simeq T_{RD} (a_{RD}/a)^{3/8}$.
At the time of decoupling $T_D > T_{RD}$ and Eq.~(\ref{eq:master})
gives
\begin{equation}
  \frac{n_B}{s} \simeq \frac{T_D^{11}}{M_\ast^2 M_P^3 T_{RD}^6}.
  \label{eq:8}
\end{equation}
This asymmetry, however, is diluted by a continuous production of
entropy.  The dilution factor is given by $(T_{RD}/T_D)^5$ and,
hence, the final asymmetry is
\begin{equation}
  \frac{n_B}{s} \simeq \frac{T_D^{6}}{M_\ast^2 M_P^3 T_{RD}}.
  \label{eq:w=0}
\end{equation}
Within the linear approximation made in Eq.~(\ref{eq:master}), the
initial asymmetry in Eq.~(\ref{eq:8}) cannot be larger than $O(1)$.
Therefore $T_{RD}$ cannot be smaller than about $10^{-2} T_D$ to
obtain the correct baryon asymmetry, which gives an upper limit
\begin{equation}
  \frac{n_B}{s} \lesssim 10^{2} \frac{T_D^{5}}{M_\ast^2 M_P^3}.
\end{equation}
This result is 3--4 orders of magnitude enhanced relative to
Eq.~(\ref{eq:w=1/3}) and allows for $T_{RD} \simeq 10^{14}$~GeV.

A third possibility is to generate the baryon asymmetry while a
non-thermal component with $w > 1/3$ dominates the universe.  The
non-thermal energy component decreases more rapidly than radiation, so
there is no need for this component to decay into additional radiation
and produce more entropy in order to enter the radiation-dominated
epoch.  Hence, once $n_B/s$ is set at $T_D$, it remains constant. The
absence of further dilution opens up the range of allowed parameters.
One example where this can occur is in the ekpyrotic
\cite{Khoury:2001wf} or cyclic \cite{cyclic} universe where the
kinetic energy density of a scalar field $\phi$ dominates immediately
after the bang over a smaller, subdominant radiation component. The
scalar field is the modulus that describes the interbrane separation.
Another possibility is that the inflaton $\phi$ falls down a steep
potential at the end of inflation and shoots out as a massless scalar
field \cite{Peebles:1998qn}. More general $w$ can be realized, for
example, by a coherent oscillation of $\phi$ about the minimum of the
potential $V(\phi) = \lambda \phi^{2N}/M_P^{2N-4}$ with a coupling
constant $\lambda$.  This form of $V(\phi)$ yields $w =(N-1)/(N+1)$,
where $1/3 < w \leq 1$ for $N > 2$ \cite{Turner:1983he}.  Such a
potential is natural in supersymmetry using a discrete symmetry, and
it is easy to verify that soft supersymmetry breaking effects do not
spoil its desired behavior.

We begin our analysis with the universe dominated by $\rho_\phi \sim
a^{-3(1+w)}$, which decreases faster than a sub-dominant radiation
component $\rho_R \sim a^{-4}$.  The $\phi$-dominated universe expands
as $a \propto t^{2/[3(1+w)]}$, while the temperature drops as $T(t) =
T_{RD} [a_{RD}/a(t)] = T_{RD} (t_{RD}/t)^{2/[3(1+w)]}$. Therefore,
\begin{eqnarray}
  \rho_\phi &\simeq& T_{RD}^4 \left( \frac{a_{RD}}{a} \right)^{3(1+w)}, \\
  \rho_R &\simeq& T_{RD}^4 \left( \frac{a_{RD}}{a} \right)^{4}.
\end{eqnarray}
Combining these relations, we find the asymmetry
\begin{equation}
  \frac{n_B}{s} \sim \frac{T_D^{8}}{M_\ast^2 M_P^3 T_{RD}^3}
  \left(\frac{T_{RD}}{T_D}\right)^{9(1 - w)/2}.
  \label{eq:enhanced}
\end{equation}
Since we have the freedom in this scenario to make $T_D$ significantly
greater than $T_{RD}$ and since $w>1/3$, the baryon asymmetry in
Eq.~(\ref{eq:enhanced}) can be significantly enhanced relative to that
in Eq.~(\ref{eq:w=1/3}) by a factor $(T_D/T_{RD})^{3(3w-1)/2}$.
Henceforth we focus on the case $w>1/3$.

\begin{figure}[t]
  \centering
  \includegraphics[width=0.93\columnwidth]{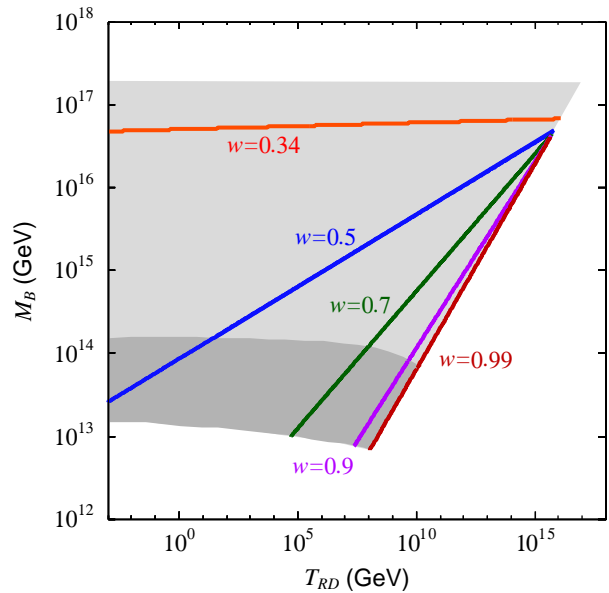}
  \caption{The range of $T_{RD}$, $M_B$ and $w$ that can generate
    an observationally acceptable baryon asymmetry assuming a
    dimension five ($n=1$) baryon number violating operator {\it and
      no significant entropy production after decoupling} is shown in
    dark gray and light gray. In supersymmetric theories, where the
    gravitino decays rapidly to an LSP with $m_{LSP} \approx 100$~GeV,
    the allowed region is restricted to the dark gray region only to
    avoid overclosure of the universe by LSP's.  Both light and dark
    gray regions are allowed if the $m_{LSP} \ll 100$~GeV, if the LSP
    decays, if the gravitino is lighter than keV, or if there is no
    supersymmetry.}
  \label{fig:D5}
\end{figure}

Next, we discuss the origin of the $B$-violating interaction that
is necessary for any of the baryogenesis scenarios considered
here.  To keep the discussion general, we assume that
$B$-violating interactions are generated by an operator ${\cal
O}_B$ of mass dimension $D = 4 + n$. The rate of such interactions
is given by $\Gamma_B = T^{2 n +
  1}/M_B^{2 n}$, where $M_B$ is the mass scale associated with ${\cal
  O}_B$.  Decoupling of $B$ violating processes occurs at $T \sim
T_D$, when $\Gamma_B$ falls below $H \sim
(T_{RD}^2/M_P)(T/T_{RD})^{3(1 + w)/2}$. The decoupling temperature
$T_D$ is then estimated to be
\begin{equation}
  T_D \sim T_{RD} \left( \frac{M_B^{2n}}{M_P \, T_{RD}^{2 n -
        1}}\right)^{{2}/{(4 n - 3 w - 1)}}.
\label{TD}
\end{equation}
When $w=1$ and $n = 1$, Eq.~(\ref{TD}) is not applicable since the
interactions with dimension five operators, such as those
responsible for the neutrino mass seesaw mechanism \cite{seesaw},
are either in thermal equilibrium all the time or never.  Using
Eqs.~(\ref{eq:enhanced}) and (\ref{TD}), we can identify the
values of $(T_{RD}, \, M_B)$ that generate $n_B/s \sim 10^{-10}$.
Assuming $n=1$,  Fig.~\ref{fig:D5} shows the result for various
values of $w$. The plots for $n>1$ are similar.

In supergravity theories, gravitino production places severe bounds on
the highest temperature $T_{max}$ attained in the early universe.
Apart from order unity coefficients, the gravitino abundance $Y_{3/2}$
is given by
\begin{equation}
  \label{eq:2}
  Y_{3/2} = \frac{n_{3/2}}{s}
  \sim 10^{-4} \frac{T_{RD}}{M_P} \left( \frac{T_{max}}{T_{RD}}
  \right)^{3(1-w)/2},
\end{equation}
where we have used Boltzmann's equation $d n_{3/2}/dt + 3H n_{3/2} =
\sigma_{\it eff} n_R^2$, with $\sigma_{\it eff} \sim 1/M_P^2$ and $n_R
\sim T^3$. We set $T_{max} = T_D$ for numerical estimates below.  The
bounds come from two constraints: ({\it i}) ensuring that the products
of BBN will not be dissociated by late gravitino decays and ({\it ii})
avoiding overclosure of the universe by gravitinos, if they are the
Lightest Supersymmetric Particles (LSP's), or by the LSP's that would
be produced in gravitino decays.  Here, we assume that the gravitino
has a mass $m_{3/2} \gtrsim 100$~TeV and decays rapidly -- before 
BBN -- as expected in anomaly mediated supersymmetry breaking scenarios
\cite{Randall:1998uk,Giudice:1998xp}.  Then, there is only constraint
$(ii)$ that the LSP's produced in gravitino decay not overclose the
universe, which requires $Y_{3/2} < 4 \times 10^{-12} ({\rm
  100~GeV}/m_{LSP})$. This range is indicated in Fig.~\ref{fig:D5} in
dark gray shading, assuming $m_{LSP} =100$~GeV.  On the other hand,
the entire range shown in Fig.~\ref{fig:D5} is allowed if $R$--parity
is violated so that the LSP decays before BBN, if the LSP is much
lighter than 100~GeV, if the gravitino is lighter than keV, or if
there is no supersymmetry at all.

It is fascinating that the correct baryon asymmetry can be obtained
without overproducing gravitinos. The typical energy scale for the $B$
violation in this scenario is about $10^{14}$~GeV as seen in
Fig.~\ref{fig:D5}. For instance, this energy scale is consistent with
what is expected for the Majorana mass of a right-handed neutrino in
the seesaw mechanism \cite{seesaw} which violates $B$ in conjunction with
electroweak anomaly effects \cite{Kuzmin:1985mm}.  This would predict
nearly degenerate Majorana neutrinos with masses above 0.1 eV and
neutrino-less double-beta decay ($0\nu \beta \beta$) at a rate
observable in near-future experiments \cite{gratta}.  In comparison,
thermal leptogenesis places a tight upper bound on the neutrino mass
$m_\nu \leq 0.11$~eV \cite{Buchmuller:2003gz} and observation of
$0\nu\beta\beta$ at larger $m_\nu$ would exclude that scenario.
Another possibility for $B$-violation is the $D=7$ operator
$W=(UDD)(UDD)/M_B^3$, which satisfies all experimental constraints
provided $M_B \gtrsim 100$~TeV.

\begin{figure}[htbp]
  \centering
  \includegraphics[width=0.93\columnwidth]{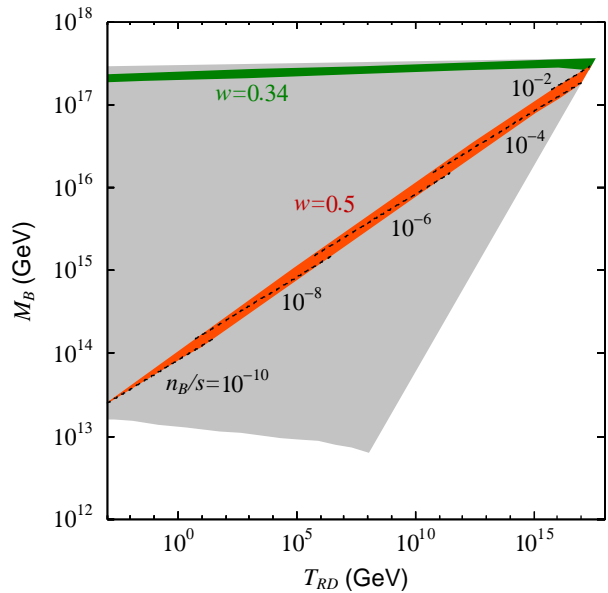}
  \caption{ The range of $T_{RD}$, $M_B$ and $w$ that can generate an
    observationally acceptable baryon asymmetry and avoid overclosure
    of the universe by stable LSP's expands to cover essentially both
    the light and dark gray regions in Fig.~\ref{fig:D5} {\it if we
      allow for entropy to be produced after the baryon-number
      violating interactions and gravitinos decouple}. For each $w$,
    the allowed line in Fig.~\ref{fig:D5} becomes an allowed strip
    (shown for two cases in the Figure) because we now include the
    possibility that the baryon asymmetry may be overproduced at
    decoupling to some degree and brought to its proper value by
    entropy production after decoupling.  For the case $w=0.5$, we
    indicate along the strip the values of $n_B/s$ before dilution.
    The additional entropy reduces the gravitino/LSP density, thereby
    opening up the allowed range of $T_{RD}$ and $M_B$.}
  \label{fig:dilution}
\end{figure}

Higher $M_B$ can be accommodated if there is significant entropy
production below $T_{RD}$.  The regions above the curves in
Fig.~\ref{fig:D5} produce a larger baryon asymmetry and we can afford
such an entropy production, which in turn also dilutes the gravitinos.
Then, a much wider region of the parameter space becomes available.
The predictions for a $D=5$ operator, with entropy production below
$T_D$, are presented in Fig.~\ref{fig:dilution}.

One additional concern in the case of inflation is that gravitinos
may be produced by quantum fluctuations in the de Sitter phase. If
the gravitino mass can be ignored during the inflation, its
coupling to the background is conformal, and no gravitinos are
produced \cite{Parker:1969au}.  Depending on the details of the
model of inflation, the gravitino mass may be enhanced and hence
its production \cite{Lemoine:1999sc} for helicity $\pm 3/2$
states.  We have checked that for certain models, {\it e.g.}\/ the
supersymmetric hybrid inflation model \cite{Dvali:ms}, the
gravitino constraint is easy to satisfy \cite{Lemoine:1999sc}.
Note that the helicity $\pm 1/2$ states are actually ``eaten''
inflatino which decays quickly and hence is harmless
\cite{Nilles:2001fg,Greene:2002ku}.  There may also be a similar
concern that gravitinos are overproduced by brane collisions in an
ekpyrotic or cyclic model, but this issue lies beyond the scope of
this Letter.

Finally, we point out that $M_\ast$ does not have to be as high as
the Planck scale.  In fact, if the $B$-violation is soft, for
example by the Majorana mass $M_R$ of the right-handed neutrino,
the operator in Eq.~(\ref{eq:operator}) does not cause any
unitarity violation up to the Planck scale even if $M_\ast^2
\simeq M_R M_P$. This is an interesting possibility that we will
not pursue further in this Letter.

In summary, we have presented a new framework for baryogenesis where
$CP$ violation lies in a gravitational interaction.  The expansion of
the universe promotes the microscopic $CP$ violation to a dynamical
violation of $CPT$ that shifts the relative energies of particles and
anti-particles.  A $CP$-conserving $B$-violating interaction in
equilibrium can then create the asymmetry that gets frozen when the
$B$-violating interaction decouples.  We have shown that it is
possible to obtain the correct magnitude of the baryon asymmetry in
many different cosmological scenarios: radiation-dominated ($w \approx
1/3$), matter-dominated ($w=0$), and kinetic-energy dominated ($1/3 <
w \leq 1$) universes.  In the last case, in particular, one can obtain
the correct baryon asymmetry while keeping the gravitino abundance low
enough to avoid overclosure by its decay products. We envision that
particle physics beyond the standard model can provide the required
$B$ violation at an energy scale above $10^{13}$~GeV, giving a whole
new class of realistic models of baryogenesis.

\begin{acknowledgements}
  We thank D.~Gross, P.~Langacker, J.~Lykken, J.~Maldacena,
  L.~Susskind, and E.~Witten for discussions.  RK is the Marvin L.
  Goldberger Member and GDK is a Frank and Peggy Taplin Member at the
  Institute for Advanced Study (IAS).  PJS is Keck Distinguished
  Visiting Professor at the IAS with support from the Wm.-Keck
  Foundation and the Monell Foundation.  HM was supported by the
  Institute for Advanced Study, funds for Natural Sciences.  This work
  was also supported in part by the DOE under contracts DE-FG02-90ER40542 and
  DE-AC03-76SF00098, and in part by NSF grant PHY-0098840.
\end{acknowledgements}


\end{document}